\documentclass[twocolumn]{autart}    % Enable this line and disable the
\usepackage[utf8]{inputenc}
\usepackage[english]{babel}

\usepackage{graphicx}
\usepackage{subfig}
\usepackage{amsmath,amssymb}
\usepackage{enumerate}

\usepackage{xcolor}
\usepackage{todonotes}

\newtheorem{lemma}{Lemma}
\newtheorem{proposition}{Proposition}

\newtheorem{remark}{Remark}
\newtheorem{assumption}{Assumption}
\newenvironment{proof}{\medskip\noindent{\it Proof. }}{ \medskip}
\newcommand{\cal}{\mathcal}

\def\begcen{\begin{center}}
\def\endcen{\end{center}}

\newcommand{\rank}{ \mbox{rank } }

\def\caly{{\cal Y}}

\def\cala{{\cal A}}

\def\hatthe{\hat{\eta}}
\def\tilthe{\tilde{\eta}}

\def\liminf{\lim_{t \to \infty}}

\def\intnum{\mathbb{Z}}

\def\hatthe{\hat{\theta}}
\def\tilthe{\tilde{\theta}}

\def\hatthe{\hat{\eta}}
\def\tilthe{\tilde{\eta}}

\def\liminf{\lim_{t \to \infty}}

\def\L2{{\cal L}_2}
\def\L2e{{\cal L}_{2e}}

\def\rea{\mathbb{R}}

\def\reapos{\mathbb{R}_{>0}}

\def\intnum{\mathbb{Z}}

\def\adj{\mbox{adj}}

\def\hatthe{\hat{\theta}}
\def\tilthe{\tilde{\theta}}

\def\begmat#1{\begin{bmatrix}#1\end{bmatrix}}
\def\begali#1{\begin{align}{#1}\end{align}}
\def\begalis#1{\begin{align*}{#1}\end{align*}}
\def\begsubequ{\begin{subequations}}
\def\endsubequ{\end{subequations}}
\def\begequarr{\begin{eqnarray}}
\def\endequarr{\end{eqnarray}}
\def\begequarrs{\begin{eqnarray*}}
\def\endequarrs{\end{eqnarray*}}
\def\begarr{\begin{array}}
\def\endarr{\end{array}}
\def\begequ{\begin{equation}}
\def\endequ{\end{equation}}
\def\lab{\label}
\def\begdes{\begin{description}}
\def\enddes{\end{description}}
\def\begenu{\begin{enumerate}}
\def\begite{\begin{itemize}}
\def\endite{\end{itemize}}
\def\endenu{\end{enumerate}}

\def\lef[{\left[\begin{array}}
\def\rig]{\end{array}\right]}

\def\begcen{\begin{center}}
\def\endcen{\end{center}}
\def\begrem{\begin{remark}\rm}
\def\endrem{\end{remark}}
\def\begassum{\begin{assumption}}
\def\endassum{\end{assumption}}
\def\begassums{\begin{assumption*}}
\def\endassums{\end{assumption*}}
\def\begassu{\begin{ass}}
\def\endassu{\end{ass}}
\def\beglem{\begin{lemma}}
\def\endlem{\end{lemma}}
\def\begcor{\begin{corollary}}
\def\endcor{\end{corollary}}
\def\begfac{\begin{fact}}
\def\endfac{\end{fact}}

\def\TAC{{\it IEEE Trans. Automatic Control}}

\def\IJC{{\it International Journal of Control}}

\def\AUT{{\it Automatica}}

\usepackage{color}

\begin{document}

\begin{frontmatter}

\title{Observability is Sufficient for the Design of Globally Exponentially Convergent State Observers for State-affine Nonlinear Systems} % Title, preferably not more

\author[UNSD]{Lei Wang}\ead{lei.wang2@sydney.edu.au},
\author[ITAM]{Romeo Ortega}\ead{romeo.ortega@itam.mx},
\author[ITMO]{Alexei Bobtsov}\ead{bobtsov@mail.itmo.ru}

%\address[zju]{College of Control Science and Engineering, Zhejiang University,  China.}
\address[UNSD]{Australia Centre for Field Robotics, The University of Sydney, Australia.}
\address[ITAM]{Departamento Acad\'emico de Sistemas Digitales, ITAM, M\'exico.}
\address[ITMO]{Faculty of Control Systems and Robotics, ITMO University,  Russia.}
%\address[zjuWREP]{Zhejiang University of Water Resources and Electric Power, Hangzhou, P.R.China.}


\begin{abstract}                          % Abstract of not more than 200 words.
In this paper we are interested in the problem of {\em state observation} of state-affine nonlinear systems.  Our main contribution is to propose a {\em globally exponentially convergent} observer that requires only the necessary assumption of {\em observability} of the system. To the best of the authors' knowledge this is the first time such a result is reported in the literature.
\end{abstract}

\end{frontmatter}

%%%%%%%%%%%%%%%%%%%%%%%%%%%%%%%%%%%%%%
%%%%%%%%%%%%%%%%%%%%%%%%%%%%%%%%%%%%%
\section{Introduction}
\lab{sec1}
%%%%%%%%%%%%%%%%
%
In this paper we are interested in the problem of state observation of  single-input single-output, {\em state-affine} nonlinear systems with dynamics of the form
\begali{
	\nonumber
	\dot x(t) & = A(u,y,t)x(t)+b(u,y,t),\;x(t_0)=x_0\in\mathbb{R}^n\\
	y(t)&=C^\top (u,t)x(t),
	\lab{nl}
}
with $x(t)\in\mathbb{R}^n$, $y(t)\in\mathbb{R}$, $u(t)\in\mathbb{R}$, which has been extensively studied in the control   literature \cite{BERbook,RUGbook}. The standard solution to this problem is given by the Kalman-Bucy filter for linear time-varying (LTV) systems. Indeed, we can write the system \eqref{nl} as an LTV system
\begali{
	\nonumber
	\dot x & = A(t)x+b(t),\;x(t_0)=x_0\\
	y&=C^\top (t)x
	\lab{ltv}
}
where, with some abuse of notation, we defined
\begalis{
	A(t)&:= A(u(t),y(t),t),\; \\ b(t)&:= b(u(t),y(t),t),\;\\ C(t)&:= C(u(t),t).
}
It is well-known that, under some reasonable boundedness assumptions, the Kalman-Bucy filter generates an {\em exponentially} convergent estimate of the system state provided the  pair $(C^\top(t),A(t))$ is {\em uniformly completely observable} (UCO). That is, if the observability Grammian of the system
\begequ
\lab{obsgra}
W(t_0,t_1):=\int_{t_0}^{t_1} \Phi^\top(\tau,t_0)C(\tau)C^\top (\tau)\Phi(\tau,t_0)d\tau
\endequ
where  $t_1 > t_0 \geq 0$ and $\Phi(t,t_0)$ is the state transition matrix of the homogeneous system $\dot x(t)=A(t)x(t)$, satisfies
\begequ
\lab{conuco}
{W(t_0,t_0+T) \geq \delta I_n,\;\forall t_0 \geq 0,}
\endequ
for some positive constants $T$ and $\delta$. See \cite[Theorem 3.3]{BERbook} for a precise formulation of this result, given in terms of the (backward) observability Grammian of the system. It is widely recognized that the UCO assumption is a {\em very restrictive} condition in many practical applications, hence the interest in designing state observers for the system \eqref{ltv} under weaker assumptions---see \cite{ARAetalifac20,BOBetalijc21,ORTetalaut,RAPDOC} for some recent contributions.

The main contribution of the paper is the proof that the necessary assumption of {\em observability} of the system $(C^\top (t),A(t))$ is sufficient to design a {\em globally exponentially convergent} observer. That is, we replace the condition \eqref{conuco} of positivity of the observability Grammian {over all intervals of length $T$} by the strictly weaker assumption of being {\em full rank} {over a finite interval from the initial time $t_0$}.  We recall that observability of the system is equivalent to the fact that the set of unobservable initial conditions
$$
\{x_0 \in \rea^n\,|\; C^\top (t)\Phi(t,t_0)x_0=0,\;\forall t \geq t_0\}
$$
is the empty set.

The developments of the paper rely on two recent contributions of the authors.
\begite
\item The use of the generalized parameter estimation-based observer (GPEBO) design technique reported in \cite{ORTetalscl,ORTetalaut}. The main novelty of GPEBO is that the state observation problem is reformulated as a problem of {\em parameter estimation} of a linear regressor equation (LRE).
\item The combination of GPEBO with the dynamic regressor extension and mixing (DREM) parameter estimator \cite{ARAetaltac,ORTetaltac} reported in \cite{WANetal}---called the  [GPEBO+DREM] (G+D) estimator. In this paper  it is shown that the parameters of a LRE can be---globally and exponentially---estimated under the weak assumption of {\em interval excitation} of the regressor, which is shown to be equivalent to {\em identifiability} of the LRE.
\endite

The interested readers are referred to the aforementioned papers for further details on GPEBO, DREM and the G+D design. Since the construction of the observer involves the application of GPEBO twice and then the use of DREM, we refer in the sequel to it as {\em 2G+D observer}.

The remainder of the paper is organized as follows. In Section \ref{sec2} we present our main result, whose proof is split into three sections. In Section \ref{sec3} we derive the LRE. Section   \ref{sec4} recalls the main result of \cite{WANetal}, that is, the G+D parameter estimator for identifiable LRE. In Section   \ref{sec5} we establish the key result that observability implies  identifiability of the LRE. We wrap up the paper with concluding remarks and some possible extensions of the main result.

\noindent {\bf Notation.} $I_n$ is the $n \times n$ identity matrix. For $x \in \rea^n$, we denote the Euclidean norm $|x|^2:=x^\top x$. All mappings are assumed smooth and all dynamical systems are assumed to be forward complete.  Given a number $n \in \intnum_{>0}$ we define the set $\bar n:=\{1,2,\dots,n\}$.
%
%%%%%%%%%%%%%%%%
\section{Main Result}
\lab{sec2}
%%%%%%%%%%%%%%%%
%
Our main result is given in the proposition below. For ease of presentation, and without loss of generality, in the sequel we will assume that the starting time for the system \eqref{ltv} is $t_0=0$. Following standard practice in observer theory \cite{BERbook} we assume that   the state trajectories of \eqref{ltv} are bounded. Moreover, we assume that the  state transition matrix of the homogeneous part of the system \eqref{ltv} verifies
\begequ
\lab{unista}
\| \Phi(t,\tau)\| \leq c_1,\;\forall t \geq \tau.
\endequ
It is well-known  \cite[Theorem 6.4]{RUGbook} that this assumption is equivalent to {\em uniform stability} of the homogeneous part of the system \eqref{ltv}. This assumption may be relaxed incorporating an output injection to achieve the latter objective---but we skip this step for brevity.

\begin{proposition}
\lab{pro1}\em
Consider the system \eqref{ltv} with $A(t),b(t)$ and $C(t)$ continuous and {\em known}. Assume the pair $(C^\top (t),A(t))$  is {\em observable}. That is, there exists a $T>0$ such that the observability Grammian \eqref{obsgra} verifies
$$
\rank\{W(0,T)\}=n.
$$
There exists a 2G+D observer whose dynamics is of the form
{
\begin{equation}
	\begin{aligned}
		\dot{\chi} & = & F(\chi,y, t)\\
		\hat x & = & H(\chi,y,t)
		\label{dynobs}
	\end{aligned}
\end{equation}
with $\chi \in \rea^{n_\chi}$, such that {\em for all} initial conditions $x(0) \in \rea^n,\;\chi(0) \in \rea^{n_\chi}$ we have
}
\begequ
\lab{obscon}
\liminf |\hat x(t)-x(t)|=0,\quad (exp),
\endequ
with all state trajectories bounded.
\end{proposition}

The proof of this result proceeds along the following steps, which are elaborated in the subsequent sections.
\begenu
\item[{\bf S1}]
Application of the GPEBO observer design technique of \cite{ORTetalaut} to derive a LRE for the system  \eqref{ltv}. Namely, we construct the bounded signals $\xi(t) \in \rea^n$ and $\Phi_A(t) \in \rea^{n \times n}$ such that the state of the  system \eqref{ltv} verifies the relation
	\begequ
	\lab{x}
	x=\xi+ \Phi_A \theta,
	\endequ
where the {\em unknown} constant vector $\theta \in \rea^{n}$ satisfies the {\em LRE}
	\begin{align}
		\lab{lre}
		\caly=\psi^\top \theta,
	\end{align}
and we defined the {\em measurable} signals  $\caly(t) \in \rea$ and $\psi(t) \in \rea^{n}$
	\begsubequ
	\lab{calypsi}
		\begali{
		\lab{caly}
		\caly&:=y - C^\top (t) \xi\\		
		\lab{psi}
		\psi&:=\Phi_A^\top C(t).
	}
	\endsubequ
\item[{\bf S2}]
Application of the G+D parameter estimator of  \cite{WANetal} that generates functions $\hat \theta(t) \in \rea^{n}$ such that
\begequ
\lab{paacon}
\liminf |\hat \theta(t)-\theta|=0,\quad (exp),
\endequ
with all estimator state trajectories bounded under the assumption of identifiablity of the LRE \eqref{lre}, that is, the existence of a set of time instants $\{t_i\}_{i\in \bar n},\;t_i \in \reapos$ such that
$$
\rank\Big\{\begmat{\psi(t_1)|\psi(t_2)|&\cdots&|\psi(t_n)}\Big\}=n.
$$

\item[{\bf S3}]
Proof that observability of the  system  \eqref{ltv} is {\em equivalent} to {\em identifiabilty} of the LRE \eqref{lre},
\item[{\bf S4}] Definition of the observed state as
\begequ
\lab{hatx}
\hat x=\xi+ \Phi_A \hat \theta,
\endequ
that, in view of \eqref{x} and \eqref{paacon}---and the boundedness of $\xi$ and $\Phi_A$---clearly verifies \eqref{obscon}.
\endenu

%
%%%%%%%%%%%%%%%%
\section{Derivation of a Linear Regression Equation for the System \eqref{ltv}}
\lab{sec3}
%%%%%%%%%%%%%%%%
%
In this section, following the GPEBO approach, we reformulate the problem of state observation as one of parameter estimation. Towards this end, we derive a LRE that is going to be used for the parameter estimation. Although the result is a particular case of  \cite[Proposition 1]{ORTetalaut}, which is applicable to a broader class of nonlinear systems and includes a coordinate change, for the sake of completeness we give the result in detail and include its proof.
\begin{lemma}
	\label{lem1}\rm
	Consider the system \eqref{ltv} and the GPEBO dynamics
	\begsubequ
	\lab{gpebodyn}
	\begali{
		\lab{dotxi}
		\dot{ \xi } & =  A(t)  \xi + b(t)\\
		\lab{dotphi}
		\dot \Phi_A &= A(t) \Phi_A,\;\Phi_A(0)=I_n.
	}
	\endsubequ
The state of the  system \eqref{ltv} verifies the relation \eqref{x} 	where the constant vector $\theta \in \rea^{n}$ satisfies the {\em LRE} \eqref{lre} with the {\em measurable} signals \eqref{calypsi}.
\end{lemma}
\begin{proof}
The error signal $e:= x-\xi$ satisfies the LTV dynamics $\dot e = A(t) e$. Now, from \eqref{dotphi} we see that $\Phi_A$ is the { {\em fundamental matrix}}\footnote{Also called principal matrix.} of the $e$ system, which satisfies \cite[Property 4.4]{RUGbook}
\begequ
\lab{phiaphi}
\Phi_A(t)=\Phi(t,0).
\endequ
Consequently, there exists a {\em constant} vector $\theta \in \rea^n$ such that
$$
e = \Phi _A\theta,
$$
namely $\theta=e(0)$. We now have the following chain of implications
\begalis{
	e = \Phi_A\theta & \Leftrightarrow \; x=\xi+ \Phi_A \theta \quad (\Leftarrow e= x-\xi) \\
	& \Rightarrow \; C^\top (t)x=C^\top (t) \xi+ C^\top (t) \Phi_A \theta  \quad (\Leftarrow C^\top (t) \times)\\
    & \Leftrightarrow \; y - C^\top (t) \xi = \psi^\top \theta \quad (\Leftarrow  \eqref{ltv},\;\eqref{psi})\\
	& \Leftrightarrow \; \caly  = \psi^\top  \theta  \quad (\Leftarrow \eqref{caly}).
}
The claim of boundedness of all signal follows from the assumption \eqref{unista} and boundedness of $x$.
\end{proof}
%
%%%%%%%%%%%%%%%%
\section{Estimation of the Parameters of \eqref{lre} via G+D}
\lab{sec4}
%%%%%%%%%%%%%%%%
%
In this section we recall the main result of  \cite{WANetal}, namely that identifiability of the LRE \eqref{lre} is sufficient to design a globally exponentially convergent parameter estimator for the LRE \eqref{lre}. Instrumental to establish this result is the proof, given in  \cite{WANetal}, that identifiability of the LRE \eqref{lre} is equivalent to  the regressor $\psi$ been {\em interval exciting} (IE) \cite{KRERIE}. That is,
\begin{equation}
\label{eq:IE}
	\int_{0}^{t_d} \psi(s) \psi^\top(s)ds \ge C_d I_n,
\end{equation}
for some  $C_d>0$ and $t_d > 0$

\begin{lemma}
	\label{lem2}\rm
Consider the LRE \eqref{lre}. Define the {\em G+D interlaced} estimator
\begsubequ
\lab{intest}
\begali{
\lab{thegk}
\dot{\hat \theta}_g & =\gamma_g \psi  (\caly -\psi^\top   \hat\theta_g),\; \hat\theta_g(0)=\theta_{g0} \in \rea^n\\
\lab{phik}
\dot { \Omega} & =    \cala(t)    \Omega ,\; \Omega(0)=I_n\\
\lab{thek}
\dot{\hat \theta}  & =\gamma \Delta  (Y -\Delta  \hat\theta),\; \hat\theta(0)=\theta_0 \in \rea^n,
}
\endsubequ
where $\gamma_g >0$, $\gamma >0$,  and we defined
\begsubequ
\lab{aydel}
\begali{
\lab{ak}
   \cala(t)&:=-\gamma_g \psi  \psi^\top  \\
\lab{dk}
   D&:=I_n-\Omega  \\
\lab{del}
\Delta  & :=\det\{D \}\\
\lab{yk}
Y  & := \adj\{D \} [\hat\theta_g -\Omega \theta_{g0}]
}
with $\adj\{\cdot \}$ denoting the adjugate matrix.
\endsubequ
If the LRE \eqref{lre} is identifiable, equivalently if $\psi $ is IE, then \eqref{paacon} holds.
\end{lemma}

 \begin{proof}
 Replacing \eqref{lre} in \eqref{thegk} yields the error dynamics for the gradient estimator
$$
\dot{\tilde \theta}_g   =\cala(t) \tilde\theta_g ,
$$
where $\tilthe_g :=\hatthe_g -\theta$, and we used the definition   \eqref{ak}.   Consequently, from  the properties of the {\em fundamental matrix} $\Omega $  defined in \eqref{phik}, we get
$$
	 \tilde \theta_g   =\Omega  \tilde\theta_g(0),
$$
which may be rewritten as
\begali{
\lab{keyide}
	  D(t)\theta &=\hat \theta_g  -\Omega  \theta_{g0}\,,
}
where we used \eqref{dk}. Multiplying \eqref{keyide} by   $\adj\{D(t)\}$ we get the following  new LRE
\begequ
\lab{ykdelk}
Y  = \Delta \theta,
\endequ
where we used \eqref{del} and \eqref{yk}. We underscore the fact that the regressor $\Delta $ is a {\em scalar}.

Replacing \eqref{ykdelk} in \eqref{thek} yields the error dynamics for each of the elements $\tilde\theta_i ,\;i \in\bar n$, of the vector  $\tilde\theta $ of the  least mean squares estimator \eqref{thek}
\begalis{
\dot{\tilde \theta}_i  & =-{ \gamma \Delta^2 }\tilde\theta_i .
}
Now, in \cite[Lemma 3]{WANetal} it is shown that  the IE assumption implies that
$$
  |\Delta(t)| =|\det\{I_n - \Omega \}|>0,
  $$
for all $t\geq t_d$, hence  $\Delta $ is PE. The proof  of exponential  convergence follows from  the well-known result \cite[Theorem 2.5.1]{SASBODbook}.
  \end{proof}
%
%%%%%%%%%%%%%%%%%%%
\section{Observability of  \eqref{ltv} is Equivalent to Identifiability of \eqref{lre}}
\lab{sec5}
%%%%%%%%%%%%%%%%%%%
%
In this section we prove that observability of the LTV system \eqref{ltv} is  {\em equivalent} to identifiability of the LRE \eqref{lre}. The interest of this condition is, obviously, that this completes the proof of our main claim, namely, that observability of the system is sufficient to design a globally exponentially convergent state observer.

\begin{proposition}
\lab{pro2}\em
The following statements are equivalent.
\begin{itemize}
  \item[(i).] The LTV system (\ref{ltv}) is observable.
  \item[(ii).] The LRE (\ref{lre}) is identifiable.
\end{itemize}
\end{proposition}

\begin{proof}
\emph{$(i)\Rightarrow(ii)$.} We prove $(i)\Rightarrow(ii)$ by contradiction. We suppose that
there exists a positive integer $n_0 <n$ such that
\begequ
\lab{ranmat}
\mbox{rank}\big[\psi(t_1)|\ldots|\,\psi(t_n)\big] \leq n_0
\endequ
holds for all time sequence $\{t_i\}_{i\in \bar n}$, with $\{\bar t_i\}_{i\in \bar n}$ being such that
\[
\mbox{rank}\big[\psi(\bar t_1)|\ldots|\,\psi(\bar t_n)\big] = n_0\,.
\]
Let $v\in\mathbb{R}^n$ be such that $|v|=1$ and
\[
\psi^\top(\bar t_i)v=0,\; \forall {i\in \bar n}.
\]
Next, we show that $\psi^\top(t)v=0$ for all $t \in \rea_{\geq 0}$ by contradiction. We suppose that there exists a $t^\sharp  \in \rea_{\geq 0}$ such that
\[
\psi^\top(t^\sharp)v\neq 0\,.
\]
This indicates that
\[
n_0 = \mbox{rank}\big[\psi(\bar t_1)|\ldots |\,\psi(\bar t_n)\big] < \mbox{rank}\big[\psi(\bar t_1)|\ldots |\,\psi(\bar t_n)|\,\psi(t^\sharp)\big]
\]
which contradicts with the assumption that \eqref{ranmat} holds for all time sequence $\{t_i\}_{i\in \bar n}$. Hence, we have $\psi^\top(t)v=0$ for all $t  \in \rea_{\geq 0}$.

With this in mind, we note that $\dot z(t)=A(t)z(t)$ by defining $z(t):=\Phi_A(t)v$, which yields
\[\begin{array}{rcl}
 v^\top W(0,t_f) v &=& \int_{0}^{t_f}|C^\top(t)\Phi_A(t)v|^2 d t \\
 &=& \int_{0}^{t_f}|\psi^\top(t)v|^2d t = 0\,
\end{array}\]
for all $t_f\geq 0$.
This clearly contradicts with the observability of the LTV system (\ref{ltv}). Therefore, there exists a time sequence $\{t_i\}_{i\in \bar n}$ such that $\mbox{rank}\big[\psi(t_1)|\ldots |\,\psi(t_n)\big] = n$, proving the statement (ii).

\medskip\noindent
\emph{$(ii)\Rightarrow(i)$.}
To prove the statement (i) we let $T> t_n$, and proceed to show that
\[
W(0,T):= \int_{0}^{T} \Phi_A^\top(t)C(t)C^\top(t)\Phi_A(t) dt >0\,.
\]
As the matrix $\big[\psi(t_1)|\ldots |\,\psi(t_n)\big]$ is full rank with identifiability, it can be seen that for any $v\in\mathbb{R}^n$ satisfying $|v|=1$, there always exists a $\bar i\in\bar n$ such that
\[
|\psi(t_{\bar i})v| >0\,.
\]
By continuity, it follows that for any $v\in\mathbb{R}^n$ satisfying $|v|=1$, there exists  an $\epsilon>0$ such that
\[
\sum_{i\in\bar n}|\psi(\hat t_{i})v|\geq |\psi(\hat t_{\bar i})v|>0\,,\; \forall \hat t_{i}\in[t_i,t_i+\epsilon]\,
\]
yielding
\[\begin{array}{rcl}
v^\top W(0,T)v &=& \int_{0}^{T} |C^\top(t)\Phi_A(t)v|^2 dt  \\
&\geq& \int_{t_{\bar i}}^{t_{\bar i}+\epsilon}|\psi^\top(t)v|^2d t >0\,.
\end{array}\]
Therefore, by recalling that such $v$ is arbitrary, it can be concluded that  $W(0,T)$ is nonsingular for $T > t_n$, proving the observability of the LTV system (\ref{ltv}).
\end{proof}
%
%%%%%%%%%%%%%%%%
\section{Concluding Remarks and Some Extensions}
\lab{sec6}
%%%%%%%%%%%%%%%%
%
We have presented in the paper a 2G+D state observer for the system \eqref{ltv} that ensures global, exponential convergence for observable systems. The  observer is of the form  \eqref{dynobs} of order  $n_\chi=(3+2n)n$ and the dynamic equations given in   \eqref{gpebodyn} and \eqref{intest} with the auxiliary signals \eqref{calypsi} and \eqref{aydel}. The observed state is computed via  \eqref{hatx}. To the best of the authors' knowledge this is the first time a result of this level of generality---concerning state-affine nonlinear systems---is reported in the literature.

The proposed 2G+D observer enjoys some {\em robustness} properties, inherited from the robustness of the G+D parameter estimator of Section \ref{sec4}. Indeed, it is shown in \cite[Proposition 4]{WANetal} that the G+D estimator defines  a bounded-input-bounded-state  operator with respect to additive disturbances to the LRE  \eqref{lre}. That is, it is shown that if
$$
\caly=\psi^\top \theta +d
$$
with bounded $d$, the state of the system remains bounded. In the present observer context such disturbances may come from noise in the state or output equations, or from uncertainty in the system matrices of \eqref{ltv}.

There are several immediate extensions of our result, in particular the case of uncertainty in the system matrices, treated in \cite{BOBetalijc21}. Also, the 2G+D observer can be easily adapted when delay measurements are present as done in   \cite{BOBetalaut21}. In those papers the parameter estimation is carried out with a standard DREM estimator, which imposes some stronger excitation constraints. Using the G+D estimator of \cite{WANetal} these requirements will clearly be relaxed.

\end{document}